# LabVIEW-based Automated Setup for Interferometric Refractive Index Probing

[1*]Nazariy Andrushchak, [2]Ivan Karbovnyk

[1] Department of Computer-Aided Design Systems, Lviv Polytechnic National University, 3 Mytropolyta Andreya str., 79013 Lviv, Ukraine

[2] Department of Electronics and Computer Technologies, Ivan Franko National University of Lviv, 107 Tarnavskogo str., Lviv 79017, Ukraine

***Corresponding author:** nandrush@gmail.com (Nazariy Andrushchak)

**Keywords:** automation, LabVIEW, refractive index, optical materials

## ABSTRACT

In the paper, we explain an automated LabVIEW controlled setup that enables interferometric measurements of refractive indices in crystalline materials using a laser light source. The setup combines a mechanical system, a microcomputer-controlled gearless drive, a Michelson interferometer, an optical detector, a data acquisition system and a LabVIEW virtual instrument for an accurate non-destructive determination of the refractive index of given plane-parallel samples.

We explain a concept, implementation, and hardware/software peculiarities of the developed system. Test experiments on different crystals yielded results that are in good agreement with available reference data. The range of potential applications of the proposed setup extends from fundamental optical research to biophotonics instrumentation, where efficient delivery of light is of crucial importance and reliable automated probing tools are needed for optical components characterization.

**INTRODUCTION**

Refractive index, *n*, is a fundamental characteristic that describes the interaction of electromagnetic waves with a material. Refractive index is a function of wavelength, and the development of devices that are designed to work in the specific range of frequencies requires accurate and non-destructive measurements of *n*.

Different methods for the measurement of *n* for transparent materials have been proposed [1-4], but experimental techniques are still being improved to increase accuracy, speed up measurement processes and extend the spectral range [5-8]. In the visible range, the most widely used methods for precise materials characterization are the ellipsometry method [10], the method of immersion interferometry and the method of minimum deflection (method of the prism) [11]. Each of these methods has its advantages and disadvantages as well as a corresponding field of application. Most of optical samples or elements for technological application are fabricated as rectangular plates, often plane-parallel, with a certain thickness in μm-mm range. The most suitable method for the refractive index measurements of such plates in the visible spectrum range is the interferometric-turning method [3, 12] that also can be extended to the millimeter-wave range [13]. Recently, heightened interest in the investigation of plane-parallel nanoporous membranes in the visible range is driven by their possible applications in biomedicine [14,15] and as active elements for electro-optics and photonics [16]. Thus, new approaches for the characterization of such materials need to be developed and tested.

It is required to have at least $10^{-3}$ accuracy of refractive index determination for most optical media to achieve competitive results in new materials characterization [17-19]. To reach this level of precision, one needs to eliminate most of the factors possibly influencing the experimental results. The automation and precise computerized control of the measuring procedure is an important step towards high accuracy and reliability of the experiment. Furthermore, an automated system powered with appropriate software can provide the capabilities of real-time control and data processing, significantly reducing characterization time. Optical setup automation can be done in different ways using specific software and a variety of external hardware to collect and analyze measured data. National Instruments LabVIEW is one of the most advanced solutions to date as it combines powerful tools for instrument control and data acquisition with a considerable set of functions for signal processing and sophisticated calculations [20, 21].

The present paper discusses the integration of LabVIEW-powered automation techniques, interferometric setup, and a gearless motor drive solution for the accurate measurements of the refractive index of crystalline materials using a laser light source.

## MATERIALS AND METHODS

*Optical setup*

The optical setup, including necessary automation elements for measuring the refractive index of crystalline materials, is shown schematically in Figure 1. The sample to be measured is placed in the test arm of the Michelson interferometer [22-24]. Light from the laser, 633 nm, is split into two beams by the semitransparent dividing prism. After splitting, one of the beams reflects from the mirror, and the other passes through the polarizer, and the sample then reflects from the other mirror and goes through the same sample once again. Both beams interfere in the semitransparent dividing prism to form the interference pattern that is focused on the Thorlabs PDA100A2 photodetector using the lens. The photodetector is equipped with an adjustable gain amplifier and the area of the pattern that can projected onto it is about 10 mm x 4 mm. The pattern is a traveling series of high and low-intensity regions, so the adjustable slit is utilized to detect only the region corresponding to a maximum or a minimum intensity of light. As the sample rotates, the output signal of the photodetector changes in a sinusoidal fashion and is recorded into PC memory via an ADC/Digital IO (Input/Output) module. This module also serves the purpose of controlling the rotation mechanism.

Figure 2 illustrates the concept of the interference pattern change registration (to measure the interference shift) when rotating the sample by a given angle. The width of the slit on the photodetector, *t,* is chosen so that only one maximum or a minimum of the interference pattern can be detected while the sample is rotated. When the maximum of the interference pattern is in the middle of the slit (see Figure 2, a), the maximum of the signal intensity corresponding to the max value is observed on the photodetector. However, if one continues to rotate the sample, the maximum of the interference pattern will be replaced by the minimum, and at the photodetector, we will observe minimal signal intensity (see Figure 2, *b*).

*Sample positioning mechanism*

A crucial element of the described setup for refractive index measurements is the precise sample positioning mechanism. We use a gearless drive (GD) based on a synchronous machine with permanent magnets (SMPM) to ensure precise control of the rotation angle. This type of drive is used in robotic manipulators, guidance systems, etc., [25] as it provides the advantages of better mechanical rigidity, higher driving power and increased reliability due to eliminated components such as couplings and bearings [25, 26].

Figure 3 shows the functional diagram of the system (left) and the actual design of the drive with its control board. Switched stator windings of the motor (M) are realized using a bridge transistor inverter (I, MOSFET type transistors IRF3205) and three-phase sinusoidal pulse width modulation (3xPWM) that is synchronized with the rotor angular position using the absolute encoder (E) from Kübler (Optic 5876).

The GD (gearless drive) system control module utilizes an inexpensive 8-bit microcontroller (Atmega128-16AU). The microcontroller is programmed to run the algorithm of control and pulse formation to drive the transistors of the inverter and DC-DC converter. The firmware contains sine-wave look-up tables, which allow creating up to 500 thousand steps for one full turn. The absolute encoder indicates a rotor position and speed (rpm). The encoder's output signal is 12-bit Gray code (4096 bits per one full turn). The subroutine that performs the conversion of encoder signals from Gray code to straight binary is included in the firmware. The binary code is then transmitted to the personal computer using the interface described below.

*Control interface*

The overall control of the experimental setup, the adjustment of various measurements parameters and data readout are realized using a personal computer via National Instruments multifunction data acquisition unit USB-6009 connected to computer's universal serial bus. Setting the direction of sample rotation is performed by sending a bit to the input port of the microcontroller through one of the USB-6009 digital output lines and the final sample position is monitored by reading encoder bits via USB-6009 digital inputs.

In total, seven digital lines were used for the bidirectional communication between a PC and the measurement setup. In this work, we developed software that defines the number of steps required to rotate the sample from the starting position to the end position. The number of steps depends upon the angle of rotation. Also, it is possible to set the speed of motor rotation in real-time while keeping the smooth, stable and desired direction of rotation of the motor (clockwise/counterclockwise). To keep track of motor positioning, the encoder's feedback data can be displayed onscreen and stored on the computer hard drive.

The signal from the photodetector is acquired through the analog channel of the USB-6009 and digitized by 14-bits ADC.

*Data processing*

Refractive index measurement experiment data processing tools were developed using National Instruments LabVIEW. The virtual instrument (LabVIEW program) allows one to monitor experimental runs, to check the rotation angle of the sample, track changes of the signal at the photodetector and perform digital signal filtering and other calculations in real-time.

**RESULTS**

*Automated refractive index measurements*

Refractive index is calculated based on the interferometric pattern shift with respect to the "zero" position in which the laser beam is strictly perpendicular to the sample surface. Figure 4 schematically illustrates the changes of the photodetector signal as the sample rotates from the initial position going through "zero" angular position along the way.

In general, "zero" position determination is a non-trivial task. The procedure applied in this work is to find an assumed "zero" position based on frequency change in the recorded signal. This assumed position is refined by calculating a refractive index based on the region between a certain angle (5-7 degrees from the possible "zero" position that is taken from the recorded data of the signal change) and this position. Ideally, if the "zero" position is correctly chosen, indices calculated for various angles should be the same (or deviate within the error of the method). That said, the algorithm of the measurement procedure Figure 4 is given by the flowchart in Figure 5. The value of the refractive index of the crystalline sample can be calculated using the formula [27]

$$n = \frac{\sin^2 \varphi \cdot n_{air}^2 + \left[(1-\cos\varphi) \cdot n_{air} - \frac{K\lambda}{2d}\right]^2}{2\left[(1-\cos\varphi) \cdot n_{air} - \frac{K\lambda}{2d}\right]}, \qquad (1)$$

where $\varphi$ defines an angle in degrees by which the sample was rotated from the initial position, $\lambda$ is the wavelength in nm of the laser used in the experiment, $d$ is the thickness in mm of the sample, $K$ denotes the interference fringe shift, and $n_{air}$ is the refractive index of the environment. The equation (1) is applicable to calculate the refractive index of crystalline materials only for parallel plate samples in the visible range. In the case of non-parallel samples, the methodology described in [28] can be used.

*Interference pattern shift analysis*

Refractive index determination using this interferometric technique requires reliable data on the shift of the interference pattern. A typical data set corresponding to interference fringes obtained when rotating 21.37 mm thick plane-parallel cut from aluminum oxide ($Al_2O_3$) crystal is shown in Figure 6 (upper part).

When we rotate the plane-parallel sample in one arm of the interferometer, it is possible to observe the oscillations of signal intensity on the photodetector and, according to previous research [29, 30], the interference fringe shift (the shift that occurs when the sample placed in the measurement channel and rotates perpendicular to the laser beam) value can be found by counting the exact number of intensity peaks (interference maxima). Whereas intensity signal fluctuations might occur while running the experiment (as is the case in the upper part of Figure 6), for accurate calculations, especially on the larger angles of rotation, where these peaks are changing very rapidly, filtering needs to be applied before further data analysis. The lower part of Figure 6 shows the result of applying for the 2nd order Butterworth infinite impulse response digital bandpass filter to the raw measured dataset. Filter specifications were adjusted to cut-off both high-frequency noise and DC component so that the filtered signal contains data centered around the zero line and without random irrelevant fluctuations. Circles indicate zero-crossing points found by the developed LabVIEW virtual instrument. Each pair of zero-crossing points corresponds to a peak counting towards a total number of interference maxima. The example in Figure 5 shows four peaks within less than two degrees range of angle. One has to note that such a narrow range of angles is useful only for demonstration purposes and a much wider range is used in actual measurements.

**RESULTS**

Crystalline samples of $Al_2O_3$ and BK7 crown glass were selected to test the automated setup. Samples were in the form of parallel plates with different thickness to see how signal processing procedures should be adjusted depending on the optical path of the beam. In both cases, the range of considered angles was set to 10 degrees.

Based on the measured data and using equation (1) the refraction index of $Al_2O_3$ at room temperature was found to be 1.774. This value is in fair agreement with the data of Malitson, Murphy, and Rodney [31] (1.765 at 644 nm and 24°C).

The second test was performed on a 10.93 mm thick plate of standard BK7 crown glass. The calculated refractive index was 1.526. The refractive index for crown glass is typically between 1.52 and 1.62. Taking into account the absolute accuracy of the investigated sample thickness determination as $\delta d = 1$ μm, uncertainty of He-Ne LHN-302 laser wavelength as $3 \cdot 10^{-8}$ μm for $\lambda$ = 0.6328 µm [32], then for the real accuracy of interference fringe shift determination $\delta k = 1$ and angle of sample rotation $\delta\varphi = 1.4 \cdot 10^{-3}$ the accuracy of the refractive index determination using interferometer-turning method is 0.005 for Sapphire ($Al_2O_3$) sample, and 0.003 for BK7 sample. The results are summarized in Table 1.

Aiming to increase accuracy, we have also calculated a refractive index for a wider angle of sample rotation starting from the initial position. For example, consideration of 20 degrees range of interferometric data leads to the results that are closer to the reference data. The index change is, in this case, is 0.002, which corresponds to the decrease of error to approximately 0.1%. Thus, a more extensive range of rotation angles slightly increases the accuracy but requires longer measurement times and a more significant number of points to be processed, making the entire experiment more time-consuming. Still, LabVIEW signal acquisition and processing capabilities provide necessary resources not only to control the experimental runs but also to process data on the fly.

**CONCLUSION**

We developed a fully automated experimental setup to accurately measure the refractive indices of parallel optical plates. The setup uses software developed in National Instrument LabVIEW. The hardware includes a high precision gearless drive based on a synchronous machine with permanent magnets to achieve a minimum rotation angle of 0.003 degrees.

All-in-one software tools developed in National Instruments LabVIEW ensure complete experimental control, data collecting, and processing from setting the rotation speed to the calculation of the refractive index.

The setup and software were tested in the measurements of the refractive index of aluminum oxide and crown glass. Comparison of obtained values with reference data verifies that the automated method produces reliable results.

The proposed technique is suitable for express laboratory or industrial use as the time needed for a single measurement is within 5-7 min.

The hardware discussed in this paper can be adapted to work with portable lasers, and the software can be ported to an embedded system, therefore making the entire experimental system more compact. The proposed technique and respective setup are intended to be most useful to the researchers in the fields of photonics and biophotonics that deal with optical characterization of plane-parallel samples for their applications.

**ACKNOWLEDGMENT**

This result of the investigation is a part of a project that has received funding from the European Union's Horizon 2020 research and innovation program under the Marie Sklodowska-Curie grant agreement No 778156. This work was also supported by the Ministry of Education and Science of Ukraine in the frame of the "SubTera" project (registration #0119U100609).

**REFERENCES**

1. Singh, S. Refractive Index Measurement and its Application. *Physica Scripta.* **2002**, 65, 167-180.

2. Coppola, G.; Ferraro, P.; Iodice, M.; et al. Method for measuring the refractive index and the thickness of transparent plates with a lateral-shear, wavelength-scanning interferometer. *Appl. Opt.* **2003**, 42, 3882-3887.

3. Gillen, G. D.; Guha, S. Refractive-index measurements of zinc germanium diphosphide at 300 and 77 K by use of a modified Michelson interferometer. *Appl. Opt.* **2004**, 43, 2054-2058.

4. Dennis, T.; Gill, E. M.; Gilbert, S. L. Interferometric measurement of refractive-index change in photosensitive glass. *Appl. Opt.* **2001**, 40, 1663-1667.

5. Andrushchak, N. A.; Karbovnyk, I. D.; Godziszewski, K.; et al. New Interference Technique for Determination of Low Loss Materials Permittivity in the Extremely High Frequency Range. *IEEE Trans. Instrum. Meas*. **2015**, 64, 3005-3012.

6. Andrushchak, N. A.; Bobitskii, Ya. V.; Maksymyuk, T. V.; et al. A new method for refractive index measurement of isotropic and anisotropic materials in millimeter and submillimeter wave range. *18th International Conference on Microwave, Radar and Wireless Communications (MIKON)*. **2010**, 273-275.

7. Smith, B.C. Fundamentals of Fourier Transform Infrared Spectroscopy. *Second Edition. CRC Press: Tayler&Francis Group*. **2011**, 207.

8. Afsar, M. N.; Ding, H. A novel open-resonator system for precise measurement of permittivity and loss-tangent. *IEEE Trans Instrum Meas.* **2001**, 50 (2), 402-405.

9. Li, X.; Wang, C.; Zhao, J.; et al. A New Method for Determining the Optical Constants of Highly Transparent Solids. *Appl. Spectrosc.* **2017**, 71(1), 70-77.

10. Adamson, P. Laser diagnostics of nanoscale dielectric films on absorbing substrate by differential reflectivity and ellipsometry. *Opt Laser Technol.* **2002**, 34 (7), 561-568.

11. Born, M.; Wolf, E. Principles of Optics. *7th ed.: Cambridge U. Press*. **2002**, 953.

12. Andrushchak, A.S. Two-beam interferometer for refractive indices measurement of the isotropic and anisotropic materials. *Patent Russia No.2102700*, **1998**.

13. Andrushchak, N. A.; Syrotynsky, O. I.; Karbovnyk, I. D.; et al. Interferometry technique for refractive index measurements at subcentimeter wavelengths. *Microwave and Opt. Technol. Lett.* **2011**, 53, 1193-1196.

14. Adiga, S.P.; Jin, C.; Curtiss, L.A.; et al. Nanoporous membranes for medical and biological applications. *Wiley Interdiscip Rev Nanomed Nanobiotechnol*. **2009**, 1(5), 568–581.

15. Santos, H. Porous Silicon for Biomedical Applications. *Woodhead Publishing*, **2014**.

16. Andrushchak, N.; Kulyk, B.; Goering, P.; et al. Study of second harmonic generation in $KDP/Al_2O_3$ crystalline nanocomposite. *Acta Phys. Pol. A*, **2018**, 133, 856-859.

17. Dmitriev, V. G.; Gurzadyan, G. G.; Nikogosyan, D. N. Handbook of Nonlinear Optical Crystals. *Springer Series in Optical Sciences, Springer Verlag, New-York*, **1997**.

18. Edwards, G. J.; Lawrence, M. A temperature-dependent dispersion equation for congruently grown lithium niobate. *Opt. Quantum Electron*. **1984**, 16, 373-375.

19. Nicholls, J. F.; Henderson, B.; Chai, B.H. Accurate determination of the indices of refraction of nonlinear optical crystals. *Appl. Opt.* **1997**, 36, 8587–8594.

20. Luna-Moreno, D.; Espinosa Sánchez, Y. M.; Ponce de León, Y.R.; et al. Campos Virtual instrumentation in LabVIEW for multiple optical characterizations on the same opto-mechanical system. *Optik.* **2015**, 126, 1923-1929.

21. Gao, P.; Wan, S.; Xiong, Y.; et al. The Design of Fiber Bragg Grating Temperature Measurement System Based on LabVIEW. *J Comput. Commun*. **2013**, 1, 27-31.

22. Ikram, M.; Hussain, G. Michelson interferometer for precision angle measurement, *Appl. Opt.* **1999**, 38, 113-120.

23. Zhang, J.; Xu, J. Q.; Gao, C. Y.; et al. Modified Michelson interferometer for probing refractive index of birefringent crystal CSBN50. *Opt Lasers Eng.* **2009**, 47, 1212-1215.

24. McKee, D. J.; Nicholls, J. F.; Ruddock, I. S. Interferometric measurement of refractive index. *European J. Phys.* **1995**, 16, 127-134.

25. Batzel, T. D.; Lee, K. Y. Commutation torque ripple minimization for permanent magnet synchronous machines with Hall effect position feedback. *IEEE Trans. Energy Convers.* **1998**, 13, 257-262.

26. Jahns, T. M.; Soong, W. L. Pulsating torque minimization techniques for permanent magnet AC motor drives – A review. *IEEE Trans. Ind. Electron.* **1996**, 43, 321-330.

27. Andrushchak, A. S.; Tybinka, B. V.; Andrushchak, N. A.; et al. Interferometric-turning method to measure the refractive index of optical materials, *Patent Ukraine No.39155*, **2009**.

28. Andrushchak, A. S.; Voronyak, T. I.; Yurkevych, O. V.; et al. Interferometric technique for controlling wedge angle and surface flatness of optical slabs. *Opt Lasers Eng.* **2013**, 51, 342-347.

29. Andrushchak, A. S.; Tybinka, B. V.; Ostrovskij, I. P.; et al. Automated interferometric technique for express analysis of the refractive indices in isotropic and anisotropic optical materials. *Opt. Lasers Eng.* **2008**, 46, 162–167.

30. Karbovnyk, I. D.; Andrushchak, N. A.; Bobitskii, Y. V. Enhanced interferometric technique for non-destructive characterization of crystalline optical materials: automated express refractive index measurements. *5th International Conference on Advanced Optoelectronics and Lasers*. **2010**, 226-227.

31. Malitson jr, I. H.; Murphy, F. V.; Rodney, W. S. Refractive Index of Synthetic Sapphire. *J. Opt. Soc. Am.* **1958**, 48, 72.

32. Mielenz, K. D.; Nefflen, K. F.; Rowley, W.R.C.; et al. Reproducibility of Helium-Neon Laser Wavelengths at 633 nm. Appl. Opt. **1968**, 7, 289-293.

33. Millodot, M. Dictionary of Optometry and Visual Science. *Butterworth–Heinemann*. **2009**, 450.

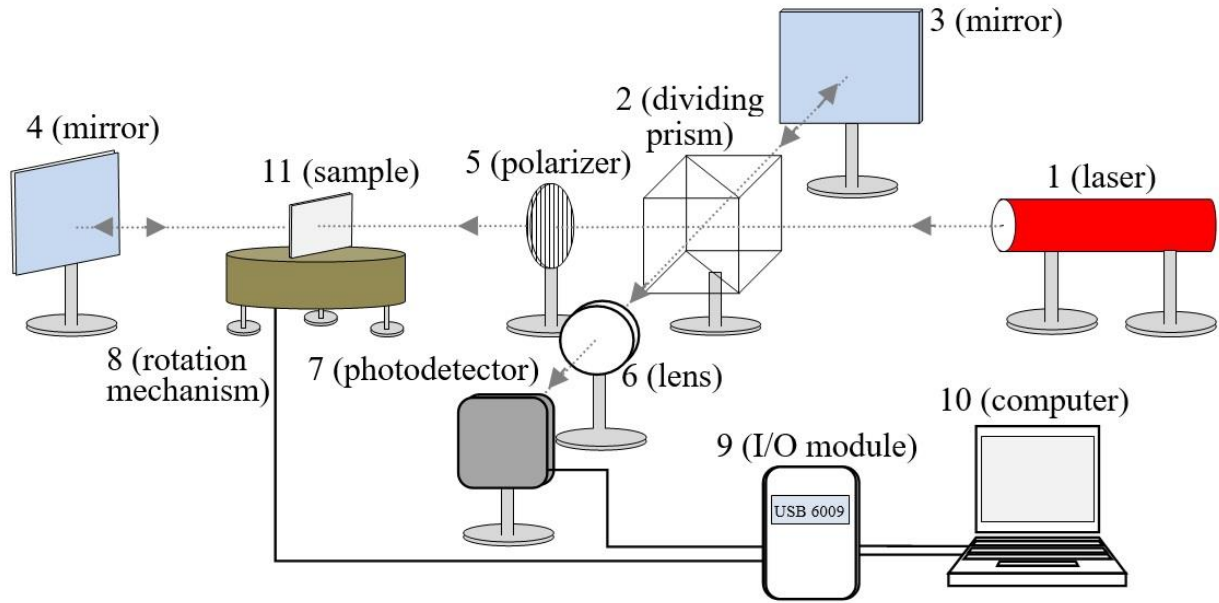

Figure 1. Experimental setup allowing automated measurements of the refractive index in the visible spectrum range. Legend: 1 – laser, 2 – semitransparent dividing prism, 3 and 4 – mirrors, 5 – polarizer, 6 –lens, 7 – photodetector, 8 – rotation mechanism, 9 – ADC and Digital I/O module (USB-6009), 10 – personal computer; 11 – sample

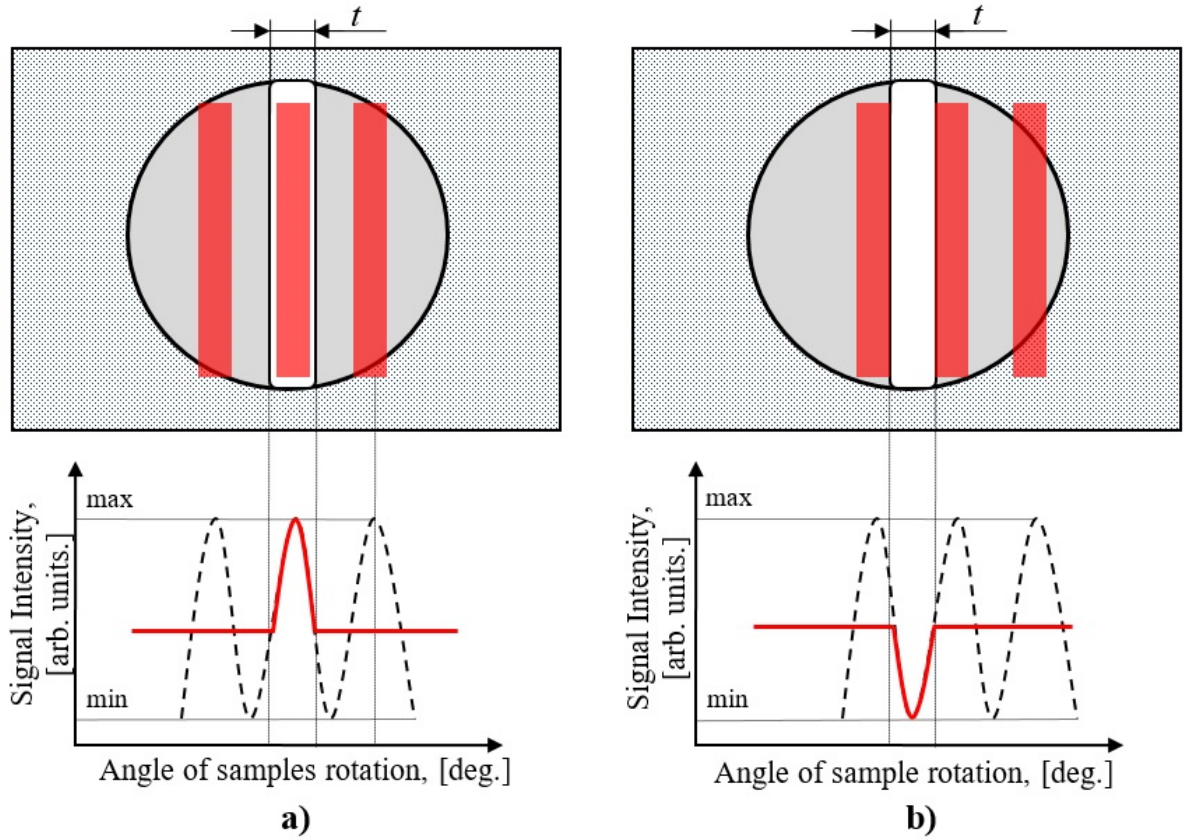

Figure 2. Change of the photodetector signal intensity: the region of the interference pattern corresponding to the maximum (a); the region of the interference pattern corresponding to the minimum (b)

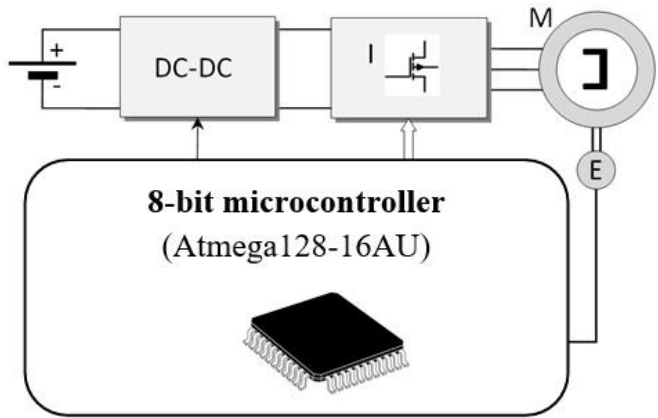 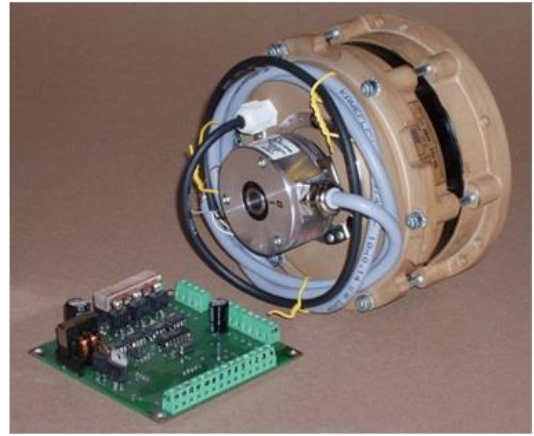

Figure 3. A functional diagram of GD left and the actual design of the GD with the control board right. Legend: M – motor, I – bridge transistor inverter, DC-DC – DC-DC converter, E – encoder

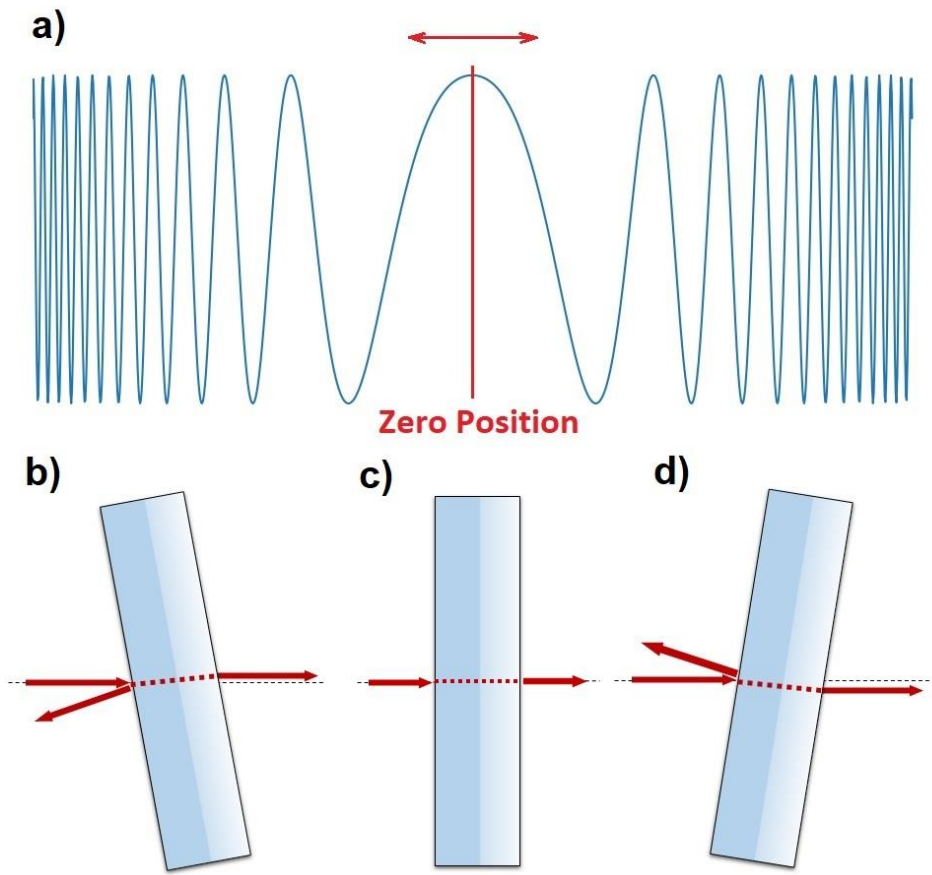

Figure 4. Schematic representation of (a) changes in photodetector signal, when (b) approaching the "zero" position in the process of clockwise rotation, (c) angular "zero" position is reached, (d) rotating clockwise from the "zero" position

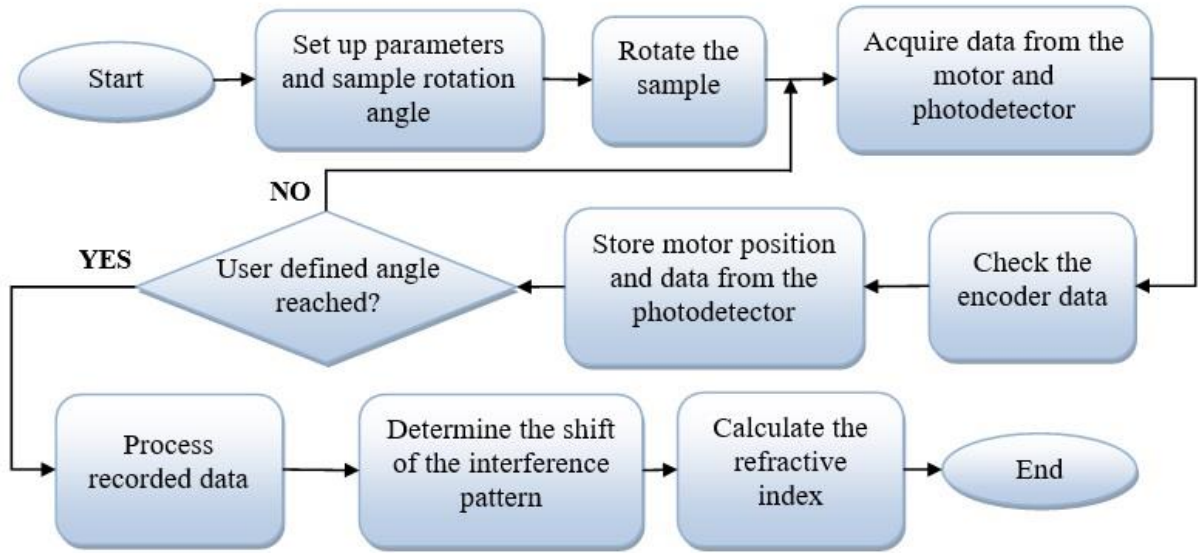

Figure 5. A flowchart explaining the process of automated refraction index measurements.

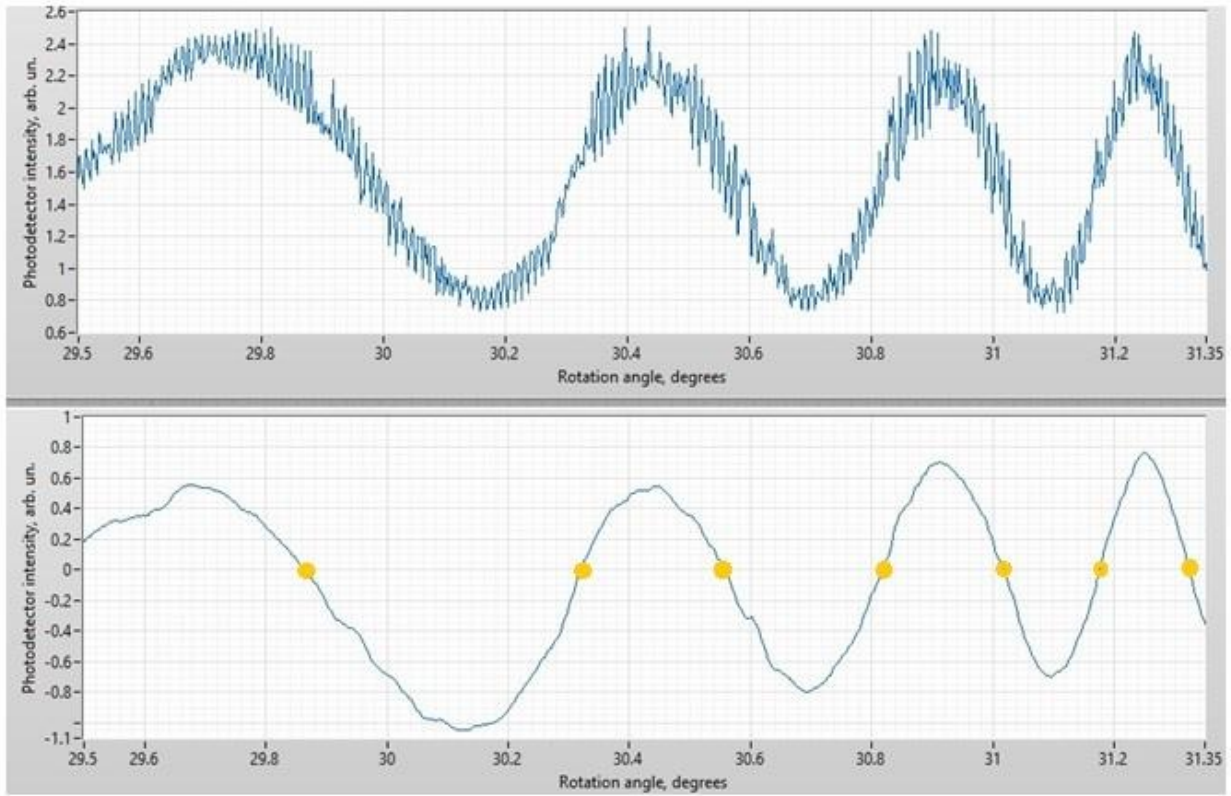

Figure 6. The typical shift of the interference fringes recorded by photodetector when rotating the plane-parallel sample from an initial position for a small angle (about 2 degrees). The upper part shows the raw (unfiltered) signal, and the bottom part shows a filtered signal.

Table 1

Comparison of measured data for the test samples using our method and data from references

| # | Sample type | Measured data | | | Reference data | | |
|---|---|---|---|---|---|---|---|
| | | Value of n | Sample thickness, mm | Experimental accuracy | Value of n | Experimental accuracy | Reference |
| 1 | $Al_2O_3$ | 1.774 | 8.16 | 0.003 | 1.7651 | 0.004 | [33] |
| 2 | BK7 crown glass | 1.526 | 10.93 | 0.005 | 1.52…1.62 | 0.007 | [33] |